\documentclass[twocolumn]{aastex631}
\usepackage{amsmath}
\usepackage{amssymb}
\usepackage{acronym}
\usepackage{hyperref}
\usepackage{color}
\usepackage{graphicx}

\usepackage{tensind}
\tensordelimiter{?}
\tensorformat{lrc}

\newcommand{\ie}{\textit{i.e.}}
\newcommand{\eg}{\textit{e.g.}}

\acrodef{ADM}{Arnowitt-Deser-Misner}
\acrodef{AMR}{adaptive mesh-refinement}
\acrodef{BH}{black hole}
\acrodef{BBH}{binary black-hole}
\acrodef{BHNS}{black-hole neutron-star}
\acrodef{BNS}{binary neutron star}
\acrodef{CCSN}{core-collapse supernova}
\acrodefplural{CCSN}[CCSNe]{core-collapse supernovae}
\acrodef{CMA}{consistent multi-fluid advection}
\acrodef{CFL}{Courant-Friedrichs-Lewy}
\acrodef{DG}{discontinuous Galerkin}
\acrodef{HMNS}{hypermassive neutron star}
\acrodef{EM}{electromagnetic}
\acrodef{ET}{Einstein Telescope}
\acrodef{EOB}{effective-one-body}
\acrodef{EOS}{equation of state}
\acrodef{FF}{fitting factor}
\acrodef{GR}{general-relativistic}
\acrodef{GRLES}{general-relativistic large-eddy simulation}
\acrodef{GRHD}{general-relativistic hydrodynamics}
\acrodef{GRMHD}{general-relativistic magnetohydrodynamics}
\acrodef{GW}{gravitational wave}
\acrodef{KHI}{Kelvin-Helmholtz instability}
\acrodef{ILES}{implicit large-eddy simulations}
\acrodef{LIA}{linear interaction analysis}
\acrodef{LES}{large-eddy simulation}
\acrodefplural{LES}[LES]{large-eddy simulations}
\acrodef{MHD}{ magnetohydrodynamics}
\acrodef{MRI}{magnetorotational instability}
\acrodef{NR}{numerical relativity}
\acrodef{NS}{neutron star}
\acrodef{PNS}{protoneutron star}
\acrodef{RMNS}{remnant massive neutron star}
\acrodef{SASI}{standing accretion shock instability}
\acrodef{SGRB}{short $\gamma$-ray burst}
\acrodef{SPH}{smoothed particle hydrodynamics}
\acrodef{SN}{supernova}
\acrodefplural{SN}[SNe]{supernovae}
\acrodef{SNR}{signal-to-noise ratio}
\acrodef{ZAMS}{zero age main sequence}

\begin{document}
 \title{Ab-Initio General-Relativistic Neutrino-Radiation Hydrodynamics Simulations of Long-Lived Neutron Star Merger Remnants to Neutrino Cooling Timescales}

\author[0000-0001-6982-1008]{David Radice}
\email{david.radice@psu.edu}
\thanks{Alfred P.~Sloan fellow}
\affiliation{Institute for Gravitation \& the Cosmos, The Pennsylvania State University, University Park, PA 16802}
\affiliation{Department of Physics, The Pennsylvania State University, University Park, PA 16802}
\affiliation{Department of Astronomy \& Astrophysics, The Pennsylvania State University,University Park, PA 16802}

\author[0000-0002-2334-0935]{Sebastiano Bernuzzi}
\affiliation{Theoretisch-Physikalisches Institut, Friedrich-Schiller-Universit{\"a}t Jena, 07743, Jena, Germany}

\begin{abstract}
  We perform the first 3D ab-initio general-relativistic
  neutrino-radiation hydrodynamics of a long-lived neutron star merger
  remnant spanning a fraction of its cooling time scale. We find that
  neutrino cooling becomes the dominant energy loss mechanism after the
  gravitational-wave dominated phase (${\sim}20\ {\rm ms}$ postmerger).
  Electron flavor anti-neutrino luminosity dominates over
  electron flavor neutrino luminosity at early times, resulting
  in a secular increase of the electron fraction in the outer layers of
  the remnant.  However, the two luminosities become comparable
  ${\sim}20{-}40\ {\rm ms}$ postmerger. A dense gas of electron
  anti-neutrinos is formed in the outer core of the remnant at
  densities ${\sim}10^{14.5}\ {\rm g}\ {\rm cm}^{-3}$, corresponding to
  temperature hot spots. The neutrinos account for ${\sim}10\%$ of the
  lepton number in this region. Despite the negative radial temperature
  gradient, the radial entropy gradient remains positive and the remnant
  is stably stratified according to the Ledoux criterion for convection.
  A massive accretion disk is formed from the material squeezed out of
  the collisional interface between the stars. The disk carries a large
  fraction of the angular momentum of the system, allowing the remnant
  massive neutron star to settle to a quasi-steady equilibrium within
  the region of possible stable rigidly rotating configurations. The
  remnant is differentially rotating, but it is stable against the
  magnetorotational instability.  Other MHD mechanisms operating on
  longer timescales are likely responsible for the removal of the
  differential rotation. Our results indicate the remnant massive
  neutron star is thus qualitatively different from a protoneutron stars
  formed in core-collapse supernovae. 
\end{abstract}

\section{Introduction}
\label{sec:introduction}
Binary \ac{NS} mergers can result in a variety of outcomes ranging from
prompt \ac{BH} formation to the formation of absolutely stable remnants,
depending on the component masses and spins and on the poorly known
\ac{EOS} of \acp{NS} \citep{Shibata:2005ss, Hotokezaka:2011dh,
Bauswein:2013jpa, Agathos:2019sah, Koppel:2019pys, Bauswein:2020aag,
Bauswein:2020xlt, Perego:2021mkd, Kashyap:2021wzs, Kolsch:2021lub,
Cokluk:2023xio}. The evolution of \ac{RMNS} that do not collapse
promptly to \ac{BH} is governed by \ac{GW} and neutrino energy and
angular momentum losses and by angular momentum redistribution due to
turbulence \citep{Radice:2020ddv, Bernuzzi:2020tgt}. \ac{GW} losses are
dominant in the first ${\sim}10{-}20\ {\rm ms}$ of the merger
\citep{Bernuzzi:2015opx, Zappa:2017xba}.  Considerable effort has been
dedicated to quantifying the \ac{GW} signal from merger remnants
\citep{Shibata:2005xz, Hotokezaka:2011dh, Bauswein:2013jpa,
Takami:2014zpa, Bernuzzi:2015rla, Bose:2017jvk, Most:2018eaw,
Breschi:2021xrx, Wijngaarden:2022sah, Breschi:2022xnc, Espino:2023llj,
Fields:2023bhs, Dhani:2023ijt}, because this is one of the key targets
for next-generation \ac{GW} experiments Cosmic Explorer
\citep{Reitze:2019iox}, Einstein Telescope \citep{Punturo:2010zz}, and
NEMO \citep{Ackley:2020atn}.  In some cases, the loss of angular
momentum support due to \ac{GW} emission can drive the \ac{RMNS} to
collapse during this phase. We refer to this outcome as a short-lived
remnant \citep{Baumgarte:1999cq, Rosswog:2001fh, Shibata:2006nm,
Baiotti:2008ra, Sekiguchi:2011zd, Palenzuela:2015dqa, Kiuchi:2022nin}.
If this does not happen, \ie, for \emph{long-lived remnants}, then the
dynamics transitions to being dominated by neutrino losses and
turbulence \citep{Hotokezaka:2013iia, Kiuchi:2017zzg, Radice:2017zta,
Radice:2018xqa, Palenzuela:2021gdo, Palenzuela:2022kqk}.

It is expected that a significant fraction of binary \ac{NS} mergers
will result in the formation of long-lived \acp{RMNS}
\citep{Piro:2017zec, Radice:2018xqa}. Such remnants have been invoked to
explain late time X-ray tails observed in a sizable fraction of short
gamma-ray bursts (SGRBs; \citealt{Dai:1998bb, Dai:1998hm, Zhang:2000wx,
Dai:2006hj, Metzger:2007cd, Rowlinson:2010a, Bucciantini:2011kx,
Rowlinson:2013ue, Metzger:2013cha, Gao:2016uwi, Geng:2016noq,
Murase:2017snw, Sarin:2022wby, Dimple:2023wvs,
Yuan:2023fqk})\acused{SGRB}, however searches of radio flares powered by
long-term energy injection of a central engine in the ejecta following
\ac{SGRB} have not yet identified a candidate \citep{Schroeder:2020qbf,
Ghosh:2022aks, Eddins:2022rlw}; see also \citet{Lu:2022wlz} for an
alternative explanation. Models of the ``blue'' kilonova in GW170817
also suggest that the remnant was stable for a timescale of at least few
tens of milliseconds \citep{Rezzolla:2017aly, Li:2018hzy,
Metzger:2018qfl, Ai:2018jtv, Nedora:2019jhl, Ciolfi:2020wfx,
Mosta:2020hlh, Combi:2023yav, Curtis:2023zfo, Kawaguchi:2023zln}.

Despite its astrophysical relevance, the evolution of long-lived \ac{NS}
merger remnants past the \ac{GW}-dominated phase of their evolution is
poorly understood \citep{Radice:2018xqa, Beniamini:2021tpy,
Margalit:2022rde}. Few \ac{NR} simulations exist of the formation of
long-lived remnants, some even spanning many tens of milliseconds
\citep{Giacomazzo:2013uua, Kastaun:2016yaf, Fujibayashi:2017puw,
Radice:2018xqa, DePietri:2019mti, Nedora:2020hxc, Ciolfi:2020wfx}.
However, current simulations either span a short timescale after the
merger \citep{Vincent:2019kor, Foucart:2020qjb, Zappa:2022rpd}, or adopt
schemes, such as the leakage scheme, that do not capture the correct
thermodynamic equilibrium of matter and neutrinos
\citep[\eg,][]{Sekiguchi:2015dma, Perego:2019adq, Hammond:2021vtv,
Palenzuela:2022kqk}.

Here, we present the first \emph{ab-initio} neutrino-radiation
hydrodynamics simulations of a long-lived \ac{NS} merger remnant
extending to more than 100~ms after the merger.  We quantify for the
first time the evolution of angular momentum in the \ac{RMNS} and in the
disk, we discuss the secular evolution of \ac{GW} and neutrino
luminosities, and the internal structure of the \ac{RMNS}.  A brief
description of the simulation and analysis methodology is given in
Sec.~\ref{sec:methods} and our results are presented in detail in
Sec.~\ref{sec:results}. Finally, we summarize and discuss our findings
in Sec.~\ref{sec:conclusions}. The appendices present additional
technical details and analysis of the simulations.

\section{Methods}
\label{sec:methods}
We consider a binary with component masses $1.35\ M_\odot$ and $1.35\
M_\odot$. We adopt the HS(DD2) \ac{EOS} (DD2 for brevity;
\citealt{Typel:2009sy, Hempel:2009mc}), which predicts a maximum
nonrotating \ac{NS} mass of $2.42\ M_\odot$ and the radius of a
nonrotating $1.4\ M_\odot$ \ac{NS} to be $R_{1.4} = 13.2\ {\rm km}$. We
construct irrotational initial data with the \texttt{Lorene}
pseudo-spectral code \citep{Gourgoulhon:2000nn}. The initial separation
is of 50~km. Evolutions are carried out using the gray moment-based
general-relativistic neutrino-radiation hydrodynamics code
\texttt{THC\_M1} \citep{Radice:2012cu, Radice:2013hxh, Radice:2013xpa,
Radice:2015nva, Radice:2021jtw}, which is based on the \texttt{Einstein
Toolkit} \citep{Loffler:2011ay, EinsteinToolkit:2021_05}. For the
simulations discussed here, we use the \texttt{Carpet} \ac{AMR} driver
\citep{Schnetter:2003rb, Reisswig:2012nc}, which implements the
Berger-Oilger scheme with refluxing \citep{Berger:1984zza,
1989JCoPh..82...64B}, and evolve the spacetime geometry using the
\texttt{CTGamma} code \citep{Pollney:2009yz, Reisswig:2013sqa}, which
solves the Z4c formulation of Einstein's equations
\citep{Bernuzzi:2009ex, Hilditch:2012fp}. 

We perform simulations using the \ac{GRLES} formalism
\citep{Radice:2017zta, Radice:2020ids} to account for angular momentum
transport due to \ac{MHD}\acused{GRMHD} effects in the remnant. Within
this formalism, turbulent viscosity is parametrized in terms of a
characteristic length scale $\ell_{\rm mix}$. We perform simulations
with $\ell_{\rm mix} = 0$ (no viscosity) and with a new density
dependent prescription for $\ell_{\rm mix}(\rho)$ calibrated using data
from high-resolution \ac{GRMHD} simulations of \citet{Kiuchi:2017zzg,
Kiuchi:2022nin}. More details are given in Appendix
\ref{sec:appendix:lmix}. The effective viscosity used in our
simulations is calibrated to high resolution simulations that span 
up to two seconds after merger, so it is expected
to be realistic. However, we cannot exclude that \ac{GRMHD} effects not
captured by the viscous prescription could also influence the dynamics.
That said, at least in the case of postmerger accretion disks around
\acp{BH} there are indications suggesting that the viscous prescription
is qualitatively and quantitatively accurate \citep{Fernandez:2018kax}.
The key conclusions of this study appear to be independent on $\ell_{\rm
mix}$, so we discuss primarily the $\ell_{\rm mix} = 0$ case in the main
text, leaving a detailed comparison of the two $\ell_{\rm mix}$
prescriptions to Appendix \ref{sec:appendix:visc}.

The evolution grid employs 7 levels of \ac{AMR}. Each model is simulated
at two resolutions with the finest grid having finest grid spacing of $h
= 0.167\ G M_\odot/c^2 \simeq 247\ {\rm m}$ and $h = 0.125\ G
M_\odot/c^2 \simeq 185\ {\rm m}$, respectively denoted as LR and SR
setups. In total, these 4 simulations required more than 10~million
CPU-core hours.

\section{Results}
\label{sec:results}

\subsection{Timescales}

Our simulations cover the last ${\sim}6$ orbits prior to merger and
extend to more than ${\sim}100$ milliseconds after merger\footnote{With
the exception of the $\ell_{\rm mix}(\rho)$ simulation at SR which was
run only up to ${\sim}64$~ms postmerger.} We
define the time of merger $t_{\rm mrg}$ as the time at which the
amplitude of the \ac{GW} $(\ell,m)=(2,2)$ mode peaks,
e.g. \citep{Bernuzzi:2014kca}. Up to merger and
shortly afterwards, the dynamics is driven by \ac{GW} losses, while
neutrinos become dominant on longer timescales. To quantify this, we
estimate the \ac{GW} and neutrino timescales as
\begin{equation}\label{eq:timescales}
  \tau_{\rm GW} = \frac{0.1\ M_\odot}{L_{\rm GW}}\,, \quad
  \tau_{\nu} = \frac{0.1\ M_\odot}{L_{\nu}}\,,
\end{equation}
where $L_{\rm GW}$ and $L_{\nu}$ are the total (all modes and all
flavors) \ac{GW} and neutrino luminosities, respectively. We have chosen
$0.1\ M_\odot$ as the typical energy scale, since this is a typical
value of the energy radiated in \acp{GW} and neutrinos after the merger
\citep{Zappa:2017xba, Radice:2018xqa}. An alternative way to estimate
the \ac{GW} luminosity is through the angular momentum as $J/\dot{J}_{\rm
GW}$ \citep{Radice:2018xqa}. This alternative method estimates values
that are broadly consistent with $\tau_{\rm GW}$.

\begin{figure}
  \includegraphics[width=\columnwidth]{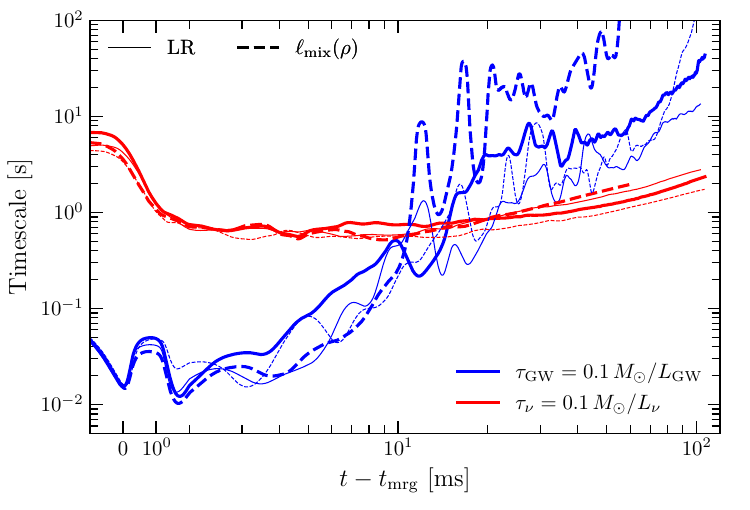}
  \caption{Gravitational wave and neutrino cooling timescales. The data
  is smoothed using a running average with window width $\Delta t = 0.5\
  {\rm ms}$. Neutrino radiation becomes the dominant mechanism for the
  evolution of the remnant ${\sim}10\ {\rm ms}$ after merger.}
  \label{fig:timescales}
\end{figure}

The timescales defined in Eq.~\eqref{eq:timescales} are shown in
Fig.~\ref{fig:timescales}. Shortly after merger $\tau_{\rm GW} \simeq
0.01\ {\rm s}$, while $\tau_{\nu} \simeq 1\ {\rm s}$.  However,
$\tau_{\rm GW}$ grows rapidly, because the remnant relaxes to a nearly
axisymmetric configuration. The neutrino luminosities decay more slowly,
so $10{-}20\ {\rm ms}$ after merger neutrinos they become the dominant
mechanism through which energy is lost by the \ac{RMNS}. At this moment,
both \ac{GW} and neutrino timescales significantly exceed the dynamical
timescale of the remnant $\tau_{\rm dyn} \sim 1\ {\rm ms}$. By
${\sim}50\ {\rm ms}$ after merger we can consider the dynamical phase of
evolution of the remnant to be concluded (see also Appendix
\ref{sec:appendix:diag}).

\subsection{Neutrino and Gravitational-Wave Luminosities}

\begin{figure}
  \includegraphics[width=\columnwidth]{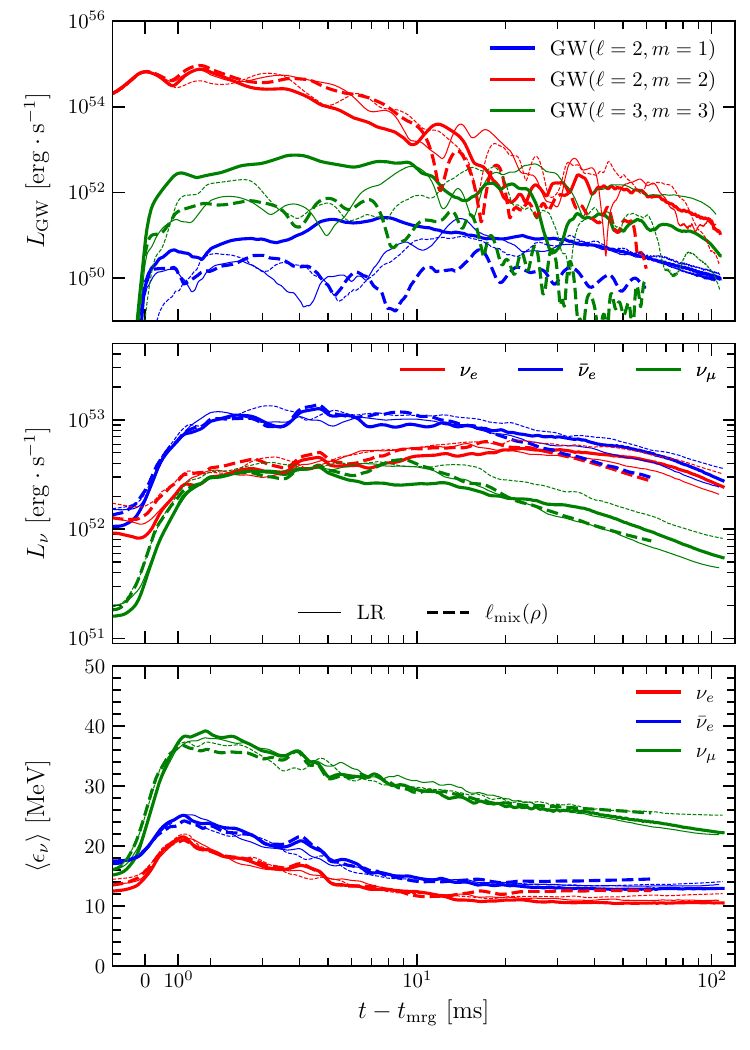}
  \caption{GW luminosity and neutrino luminosity and average energies.
  The data is smoothed using a running average with window width $\Delta
  t = 0.5\ {\rm ms}$. The GW luminosity is shown for the  three most
  dominant modes: $(\ell,m)=(2,1), (2,2)$, and $(3,3)$. $L_{\nu_\mu}$ is
  the luminosity of one of the heavy-lepton neutrino species. We do not
  track separately $\nu_\mu$, $\nu_\tau$ and their antiparticles, so the
  overall neutrino luminosity of the four species combined is $4\times
  L_{\nu_\mu}$.}
  \label{fig:luminosities}
\end{figure}

Figure~\ref{fig:luminosities} shows the \ac{GW} luminosity for the three
dominant modes, $(\ell,m)=(2,1), (2,2)$, and $(3,3)$, together with neutrino
luminosities and average energies. The \ac{GW} luminosity peaks at
merger and then decays rapidly, particularly for the simulations with
$\ell_{\rm mix}(\rho)$. The $(\ell,m)=(2,2)$ mode remains the largest
throughout the evolution, even though it is initially decaying more
rapidly than the other two. At late times, its luminosity becomes
comparable to those of the $(\ell,m)=(3,3)$ and $(\ell,m)=(2,1)$ modes.

The $\bar{\nu}_e$ luminosity is the largest in the first ${\sim}20\ {\rm
ms}$ of the merger. This is due to the fact that the beta-equilibrium
shifts to higher $Y_e$ at the extreme densities attained in the
postmerger \citep{Cusinato:2021zin}. As a result, we observe the
electron fraction close to the surface of the \ac{RMNS}, which we define
to be $\rho = 10^{13}\ {\rm g}\ {\rm cm}^{-3}$, to grow from
${\sim}0.05$ to ${\sim}0.07$ within the first ${\sim}100\ {\rm ms}$ of
postmerger evolution (Fig.~\ref{fig:profiles_xy}). The $Y_e$ grows to
$0.09$ at the inner edge of the accretion disk. Eventually, at $t -
t_{\rm mrg} \simeq 20{-}40\ {\rm ms}$, the $\nu_e$ luminosity becomes
comparable to the $\bar{\nu}_e$ luminosity.  The average neutrino
energies always satisfy $\langle \epsilon_{\nu_e} \rangle < \langle
\epsilon_{\bar\nu_e} \rangle < \langle \epsilon_{\nu_\mu} \rangle$, as a
result of the fact that the $\nu_e$'s decouple from matter at lower
densities and temperatures compared to the other two species
\citep{Endrizzi:2019trv}.

\subsection{Internal Stratification of the RMNS}

\begin{figure*}
  \includegraphics[width=\textwidth]{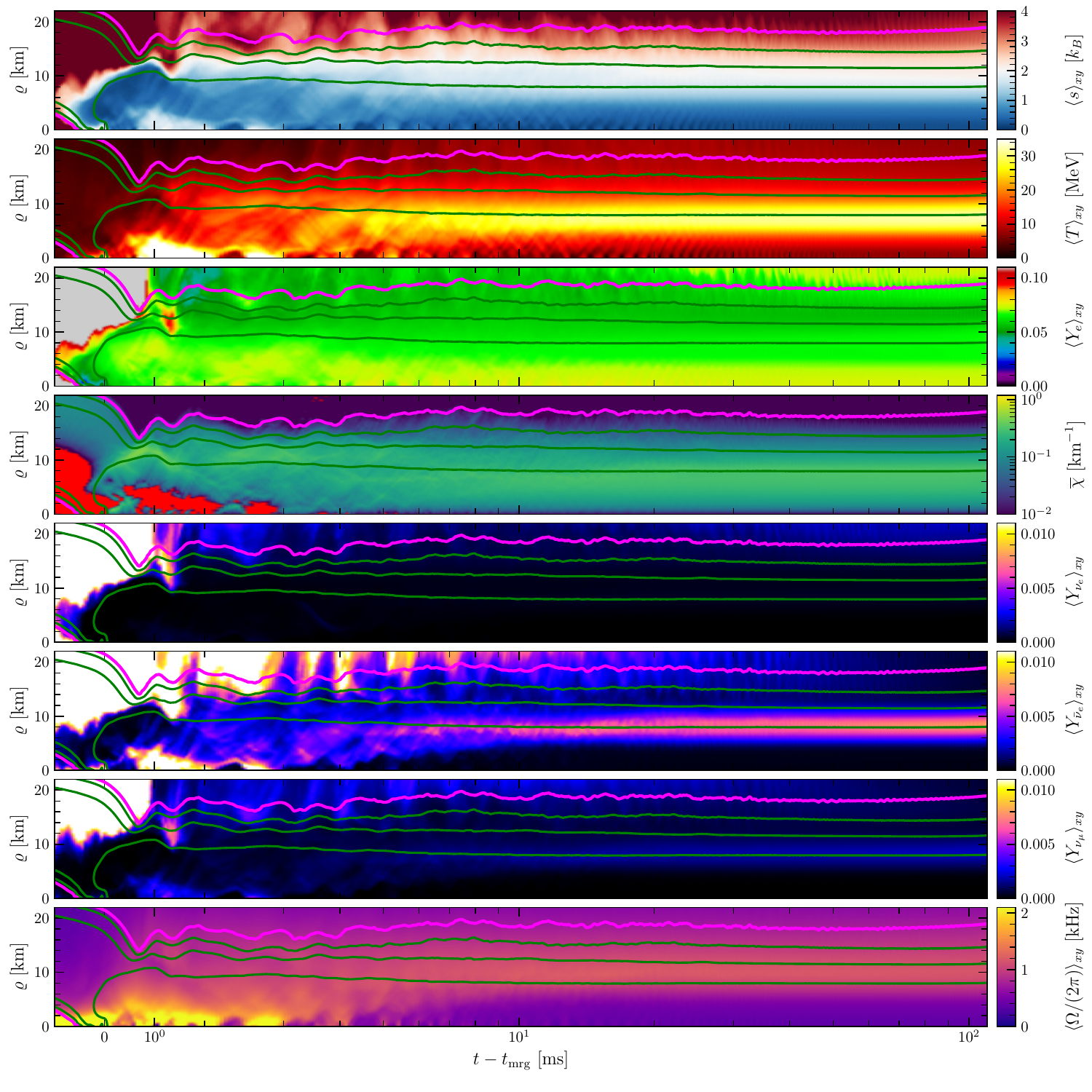}
  \caption{Angularly averaged profiles of entropy per baryon $s$,
  electron fraction $Y_e$, convection instability criterion $\chi$,
  neutrino fractions $Y_{\nu_e}$, $Y_{\bar{\nu}_e}$, and $Y_{\nu_\mu}$,
  and angular frequency $\Omega$ on the equatorial plane for the
  $\ell_{\rm mix} = 0$ SR binary. The profiles are shown as a function
  of cylindrical radius $\varrho$ and time from merger $t - t_{\rm
  mrg}$.  The purple contour denotes $\rho = 10^{13}\ {\rm g}\cdot {\rm
  cm}^{-3}$, while the green contours are $\rho = 10^{13.5}, 10^{14}$,
  and $10^{14.5}\ {\rm g}\cdot {\rm cm}^{-3}$.}
  \label{fig:profiles_xy}
\end{figure*}

Figure~\ref{fig:profiles_xy} shows azimuthally-averaged space and time
diagrams (as a function of time and cylindrical radius) for various
thermodynamic quantities on the equatorial plane for the $\ell_{\rm mix}
= 0$ SR simulation. Appendix~\ref{sec:appendix:vert} discusses a similar
analysis performed along the $z$-axis, while
Appendix~\ref{sec:appendix:visc} discusses the results for the
$\ell_{\rm mix}(\rho)$ simulation. We find that the material heated up
to high temperatures during the merger settles in a hot annular region
with $\rho \simeq 10^{14.5}\ {\rm g}\ {\rm cm}^{-3}$, as also previously
found by \citet{Bernuzzi:2015opx} and \citet{Kastaun:2016yaf}.  This
region is rich of $\bar\nu_e$, which contribute ${\sim}10\%$ of the
overall lepton number in this region. The inner core and the mantle of
the \ac{RMNS} remain at lower temperatures, creating a negative radial
temperature gradient.  Nevertheless, the radial entropy gradient is
positive at all times, stabilizing the star against convection and we
find no evidences of radial overturn in the \ac{RMNS}.

To quantify the stability to convection we compute
\begin{equation}\label{eq:chi}
  \bar\chi = \frac{1}{\rho_0} \left[ \left( \frac{\partial \rho}{\partial p} \right)_{s, Y_e}
  \frac{\partial \langle p \rangle_{xy}}{\partial \varrho} - \frac{\partial
  \langle \rho \rangle_{xy}}{\partial \varrho} \right]\,,
\end{equation}
where $\rho$ is the rest-mass density, $p$ is the pressure, $s$ is the
entropy, $Y_e$ is the proton fraction, and $\varrho$ is the cylindrical
radius. The first term in parenthesis in Eq.~\eqref{eq:chi} is obtained
from the \ac{EOS}. The factor $\rho_0 = 2.7\times 10^14\ {\rm g}\ {\rm
cm}^{-3}$ is introduced so that $\bar\chi$ has dimension of an inverse
length scale.  If $\bar\chi > 0$, then fluid elements displaced
adiabatically upwards are more dense than their environment and will not
become buoyant. In other words, the fluid is stable against convection
if $\bar\chi > 0$, this is the so-called Ledoux condition for
instability\footnote{This condition is derived in Newtonian theory, but
it is also valid in general relativity, since it is local and the
equivalence principle holds.}. $\bar\chi$, also shown in
Fig.~\ref{fig:profiles_xy}, is always positive, with the exception of a
few regions around the time of merger (highlighted in red). However, we
remark that the Ledoux instability condition only applies when the
background flow is stationary, so it is not applicable during the most
dynamical phase of evolution around $t \simeq t_{\rm mrg}$. The angular
velocity increases radially within the star and peaks at the surface, as
also found in \citet{Kastaun:2014fna} and \citet{Hanauske:2016gia}. This
implies that the remnant is not only stable against convection, but also
against the \ac{MRI}.

\subsection{Angular Momentum Balance}

\begin{figure}
  \includegraphics[width=\columnwidth]{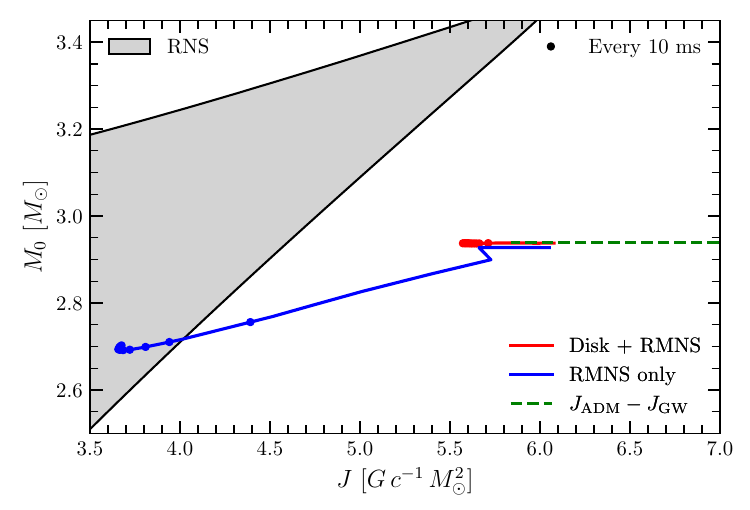}
  \caption{Baryonic mass and angular momentum at the end of the
  simulations for the \texttt{SR} binary. The grey shaded area shows the
  set of all rigidly-rotating equilibrium configurations. The green line
  shows the evolution of the angular momentum due to gravitational wave
  emission, while the red and blue line show the evolution of the
  baryonic mass and angular momentum for the disk and the remnant,
  respectively. The remnant settles to a region corresponding to a
  stable rigidly-rotating configuration, but it is still rotating
  differentially at the end of our simulations}
  \label{fig:phase_space}
\end{figure}

Figure~\ref{fig:phase_space} shows the evolution of the $\ell_{\rm
mix}=0$ SR binary in the total angular momentum $J$ and baryonic mass
$M_0$ plane (see Appendix \ref{sec:appendix:visc} for the same analysis
in the $\ell_{\rm mix}(\rho)$ case). We estimate the angular momentum of
the binary in two ways. First, as $J_{\rm ADM} - J_{\rm GW}$ (green
dashed line in the figure), where $J_{\rm ADM}$ is the ADM angular
momentum of the system (which is constant in time), and $J_{\rm GW}$ is
the angular momentum carried away by \acp{GW}. Second, as the volume
integral of the stress energy tensor (red line), $T_{\mu\nu} n^\mu
\phi^\nu$ where $n^\mu$ is the normal to the $t = {\rm const}$
hypersurface and $\phi^\mu$ is the generator of the rotations in the
orbital plane, in a cylinder of (coordinate) radius $R = 300\, G/c^2\,
M_\odot \simeq 443\ {\rm km}$. This second estimate is strictly only
valid when the spacetime is stationary and axisymmetric, as such the
dynamics close to merger, when a kink appear in the angular momentum of
the \ac{RMNS} might not be physically meaningful.  Nevertheless, the two
estimates agree well even at merger. At later times, angular momentum
evolution is determined by mass ejection and disk formation, while
$\dot{J}_{\rm GW} \simeq 0$. The shaded region in
Fig.~\ref{fig:phase_space} shows the allowed range for rigidly rotating
equilibrium configurations. We remark that $R = 300\, G/c^2\, M_\odot$
is the approximate location at which the energy due to nuclear
recombination of the disk material is approximately equal to the
gravitational potential energy, so matter can be considered to be
unbound past this point \citep{Radice:2018xqa}.

We see that binary moves on an horizontal line during the inspiral and
through merger, as \ac{GW} carry away angular momentum, but not rest
mass. Angular momentum loss is dominated by \ac{GW} emission for
${\sim}20\ {\rm ms}$ after the merger (see also
Fig.~\ref{fig:timescales}). Within this time, the two estimates for the
angular momentum agree well. On longer timescales angular momentum and
total mass are carried away by neutrino- and spiral-wave-driven outflows
\citep{Nedora:2019jhl, Nedora:2020hxc}, so the \ac{GW} estimate is no
longer reliable and the red line in Fig.~\ref{fig:phase_space} provides
the correct representation of the state of the system.

The blue line in Fig.~\ref{fig:phase_space} shows the angular momentum
and rest-mass pertaining to the inner core of the \ac{RMNS} ($\rho >
10^{13}\ {\rm g}\ {\rm cm}^{-3}$). The difference between the blue and
the red line is the total mass and angular momentum of the disk. The
latter is formed of material squeezed out of the collisional interface
between the two cores of the \acp{NS} \citep{Radice:2018pdn,
Zenati:2023lwh}. The mass of the disk saturates within ${\sim}20\ {\rm
ms}$ of the merger at $0.25\ M_\odot$. The disk contains a substantial
fraction of the angular momentum of the remnant, allowing the \ac{RMNS}
to settle to a quasi-steady state within the range of allowed
equilibria. We remark that the \ac{RMNS} is still differentially
rotating at this stage (Fig.~\ref{fig:profiles_xy}), so the ultimate
outcome of the evolution of this binary will depend on the interplay
between accretion, driven by \ac{MHD} torques, and dissipation of the
differential rotation within the remnant \citep{Radice:2018xqa,
Just:2023wtj}.

\section{Discussion}
\label{sec:conclusions}
We have performed long-term, neutrino-radiation \ac{GRHD} simulations of
a binary \ac{NS} merger producing a long-lived \ac{RMNS}.
Simulations were performed at two grid resolutions and
both with and without a treatment for subgrid scale angular momentum
transport due to turbulent viscosity. While there are some quantitative
differences between the results found in our simulations, the trends
discussed here are robust.

During the inspiral and in the first ${\sim}10{-}20$~ms after merger,
\ac{GW} emission drives the dynamics of the system. However, neutrino
radiation becomes the dominant energy-loss mechanism ${\sim}20 \ {\rm
ms}$ after merger. We find no evidence for a revival of the \ac{GW}
signal due to convective instabilities. In contrast, we show that the
remnant is stable against convection, in tension with the claims of
\citet{DePietri:2019mti}. The \ac{GW} luminosity is sensitive to
resolutions and viscosity, while neutrino radiation is robustly captured
in our simulations, within the approximations inherent to the gray-M1
scheme we are using. This highlights the importance of neutrinos and
resolution for quantitative predictions of the postmerger dynamics, as
anticipated by \citet{Zappa:2022rpd}.

Material that has been heated up by stirring during the merger forms a
hot annular region in the outer core of the \ac{RMNS}, at densities
$\rho \simeq 10^{14.5}\ {\rm g}\ {\rm cm}^{-3}$. Our simulations did not
include \ac{MHD}-effects explicitly, other than through the \ac{GRLES}
formalism. However, this is the region in which strong, small-scale
magnetic fields generated by the \ac{KHI} are expected to settle in the
aftermath of the merger \citep{Kiuchi:2014hja, Combi:2023yav}. Along the
rotational axis, this corresponds to a depth of 1~km below the surface.
If \ac{RMNS} are a viable central engine for \ac{SGRB}, then the field
need somehow to bubble out of the remnant and form large scale magnetic
structures \citep{Mosta:2020hlh}. Recent \ac{GRMHD} simulations suggest
that it is possible for the field to break out of the star
\citep{Most:2023sft, Combi:2023yav}. However, our simulations indicate
that the \ac{RMNS} is stably stratified, so it remains unclear how the
magnetic fields can emerge from it. On the other hand, the simulations
of \citet{Most:2023sft} and \citet{Combi:2023yav} used grid resolutions
of 250~m and 180~m, respectively, which are likely insufficient to
resolve the dynamics in this layer. Understanding whether
or not, and on which timescale the magnetic fields amplified by the
\ac{KHI} can emerge from the \ac{RMNS} will require high-resolution
\ac{GRMHD} simulations with realistic microphysics.

Trapped neutrinos also play an important role in the dynamics of
material in the hot layers, since they account for about $10\%$ of the
lepton number. The anti-neutrinos are formed as the result of the
conversion of degenerate neutrons into protons. This process relieves
some of the neutron degeneracy and results in an overall decrease of the
pressure of matter \citep{Perego:2019adq}. Such effect has been
neglected in previous simulations that use leakage-based approaches,
which assume nearly-frozen composition at high-optical depths. However,
it might have implications for the stability of remnants closer to the
BH formation threshold \citep{Radice:2018xqa, Zappa:2022rpd}. 

A massive accretion disk is formed by the ejection of material squeezed
out of the collisional interface between the two stars forming a massive
disk in the first ${\sim}20\ {\rm ms}$ after merger. The disk carries a
significant fraction of the angular momentum of the binary. As the disk
forms, the inner part of the remnant settles to a quasi-steady state
that corresponds to a stable uniformly rotating equilibrium
configuration. However, differential rotation persists throughout our
simulations, even when we include a treatment of subgrid angular
momentum transport calibrated against very high-resolution \ac{GRMHD}
simulations. Given that the remnant is stable against both convection
and the \ac{MRI}, it is not clear what physical mechanism will operate
to bring it to solid body rotation. This could occur as a result of
magnetic winding \citep{Baumgarte:1999cq}, possible doubly diffusive
instabilities \citep{1996ApJ...458L..71B}, or other \ac{GRMHD}
instabilities, such as the Tayler \citep{1973MNRAS.161..365T,
1973MNRAS.163...77M, 1973MNRAS.162..339W, Lasky:2011un, Ciolfi:2011xa,
Sur:2021awe}, or Tayler-Spruit dynamo \citep{2019MNRAS.485.3661F,
Margalit:2022rde}. Addressing this question will the object of future
work.

\begin{acknowledgements}
We thank Brian Metzger for helpful comments on on an early draft of our
manuscript.
DR acknowledges funding from the U.S. Department of Energy, Office of
Science, Division of Nuclear Physics under Award Number(s) DE-SC0021177
and from the National Science Foundation under Grants No. PHY-2011725,
PHY-2020275, PHY-2116686, and AST-2108467.
This research used resources of the National Energy Research Scientific
Computing Center, a DOE Office of Science User Facility supported by the
Office of Science of the U.S.~Department of Energy under Contract
No.~DE-AC02-05CH11231.
SB knowledges funding from the EU Horizon under ERC Consolidator Grant,
no. InspiReM-101043372 and from the Deutsche Forschungsgemeinschaft, DFG,
project MEMI number BE 6301/2-1.
The authors acknowledge the Gauss Centre for Supercomputing
e.V. (\url{www.gauss-centre.eu}) for funding this project by providing
computing time on the GCS Supercomputer SuperMUC-NG at LRZ
(allocation {\tt pn36ge} and {\tt pn36jo}).
\end{acknowledgements}

\appendix

\section{Mixing length prescription}
\label{sec:appendix:lmix}

Within the \ac{GRLES} formalism, angular momentum transport in the
remnant is described as an effective turbulent subgrid stress tensor
\citep{Radice:2017zta, Radice:2020ids}:
\[
  \tau_{ij} = - 2 \nu_T (e + p) W^2 \left[ \frac{1}{2} \Big( D_{i}
  v_{j} + D_{j} v_i \Big) -
  \frac{1}{3} D_k v^k \gamma_{ij} \right]\,.
\]
where $D_i$ is the covariant derivative consistent with the spatial
metric $\gamma_{ij}$, $e$ is the energy density, $p$, is the pressure,
$v^i$ is the velocity, $W$ is the Lorentz factor, and $\nu_T$ is
turbulent viscosity, which we write in terms of the mixing length
$\ell_{\rm mix}$ and the sound speed $c_s$ as
\[
  \nu_T = \ell_{\rm mix}\, c_s\,.
\]
See \citet{Duez:2020lgq} for a discussion of possible alternative
formulations.  In the context of accretion disk theory, turbulent
viscosity is typically parametrized in terms of a dimensionless constant
$\alpha$ linked to $\ell_{\rm mix}$ through the relation $\ell_{\rm mix}
= \alpha\, c_s\, \Omega^{-1}$, where $\Omega$ is the angular velocity of
the fluid \citep{1973A&A....24..337S, 1981ARA&A..19..137P}.

\begin{figure}[h]
  \centering
  \includegraphics[width=0.5\columnwidth]{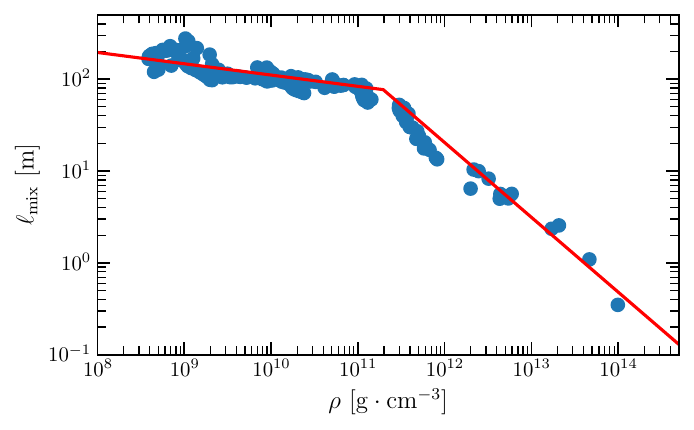}
  \caption{Mixing length prescription used for the viscous simulations.
  The data points are obtained by combining the values of $\alpha$
  reported by \citet{Kiuchi:2017zzg, Kiuchi:2022nin} with
  angular velocities and sound speed values of \citet{Radice:2020ids}.
  The solid line is the fitting function adopted here.}
  \label{fig:mixing_length}
\end{figure}

\begin{table}[h]
  \caption{Values of the fitting coefficients used in Eq.~\ref{eq:lmix}
  in units in which $M_\odot = G = c = 1$.}
  \begin{center}
  \begin{tabular}{ll}
    \tableline
    $\xi_0$ & $-10$ \\
    $\xi_1$ & $-6.5$ \\
    $a$      & $-2.0779231883396205$ \\
    $b$      & $-0.1225572350860868$ \\
    $c$      & $-0.8140456022634058$ \\ 
    \tableline
  \end{tabular}
  \end{center}
  \label{tab:mixing_length}
\end{table}

To estimate $\ell_{\rm mix}$ we take the values of $\alpha$ as a
function of rest mass density $\rho$ reported by \citet{Kiuchi:2017zzg,
Kiuchi:2022nin} and values of $\Omega$ and $c_s$ from
\citet{Radice:2020ids}. These values are reported in
Fig.~\ref{fig:mixing_length}. 
We construct piecewise linear fit for
$\ell_{\rm mix}$ in the form
\begin{equation}
\label{eq:lmix}
  \log_{10} \ell_{\rm mix}(\xi) = \left\{
  \begin{array}{ll}
    a + b\, \xi_0\,, & \xi \leq \xi_0\,, \\
    a + b\, \xi\,, & \xi_0 < \xi \leq \xi_1\,, \\
    a + b\, \xi_0 + c\, (\xi - \xi_0)\,, & \xi > \xi_1\,,
  \end{array}
  \right.
\end{equation}
where $\xi = \log_{10} \rho$. The fitting coefficients are given in
Tab.~\ref{tab:mixing_length} and the results of the fit are shown in
Fig.~\ref{fig:mixing_length}. Note that, because the piecewise linear
fit would predict divergent $\ell_{\rm mix}$ at low densities, we need
to limit the value of $\ell_{\rm mix}$ at low densities $\rho <
10^{\xi_0} = 10^{-10}\ c^6 G^{-3} M_\odot^{-2} \simeq 6 \times 10^7\
{\rm g}\ {\rm cm}^{-3}$, as shown in Eq.~\eqref{eq:lmix}.

We remark that the simulations of \citet{Kiuchi:2017zzg} employed
extremely high resolution of $\Delta x = 12.5\ {\rm m}$ and extended for
over 30~ms. They also employed a very large initial magnetic field
(${\sim}10^{15}\ {\rm G}$). As such they provide a realistic upper limit
for the viscosity inside the remnant. The more recent simulations by
\citet{Kiuchi:2022nin} not only employ very high resolution, but also
extend for over two seconds. As such, they provide realistic estimates
for the viscosity also in the disk. Moreover, \citet{Fernandez:2018kax}
found excellent agreement between \ac{GRMHD} and alpha-viscosity
simulations of postmerger accretion disks around \acp{BH}. As such, we
expect that realistic merger evolution should lay in between the
predictions of our simulations with and without \ac{GRLES}.

\section{Diagnostics}
\label{sec:appendix:diag}

\begin{figure}
  \centering
  \includegraphics[width=0.5\columnwidth]{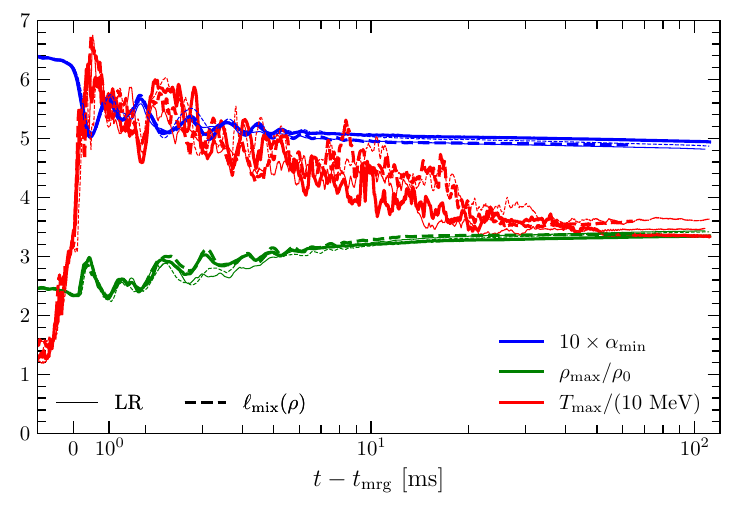}
  \caption{Time evolution of diagnostic quantities in the simulations:
  minimum of the lapse function, $\alpha_{\min}$, maximum rest-mass
  density, $\rho_{\max}$, and maximum temperature $T_{\max}$. The
  dynamical phase of evolution terminates ${\sim}50\ {\rm ms}$ after
  merger.}
  \label{fig:diagnostics}
\end{figure}

Figure~\ref{fig:diagnostics} shows the time evolution of the minimum of
the lapse function, the maximum rest-mass density, and the maximum
temperature in our simulations. The remnant is heated to temperatures of
up to $70\ {\rm MeV}$ during merger. As material initially in the ``hot
spots'' formed during merger is mixed with the rest of the \ac{NS}
matter the maximum temperature decreases \citep{Bernuzzi:2015opx,
Kastaun:2016yaf}. The \ac{RMNS} also undergoes violent oscillations in
the first milliseconds of its formation, seen in both the maximum
density and central lapse values. However, the most dynamical phase
terminates ${\sim}50\ {\rm ms}$ after the merger, after which the
oscillations have subsided and the temperature stops evolving on a
dynamical timescale.

\section{Vertical Stratification}
\label{sec:appendix:vert}

\begin{figure*}
  \includegraphics[width=\textwidth]{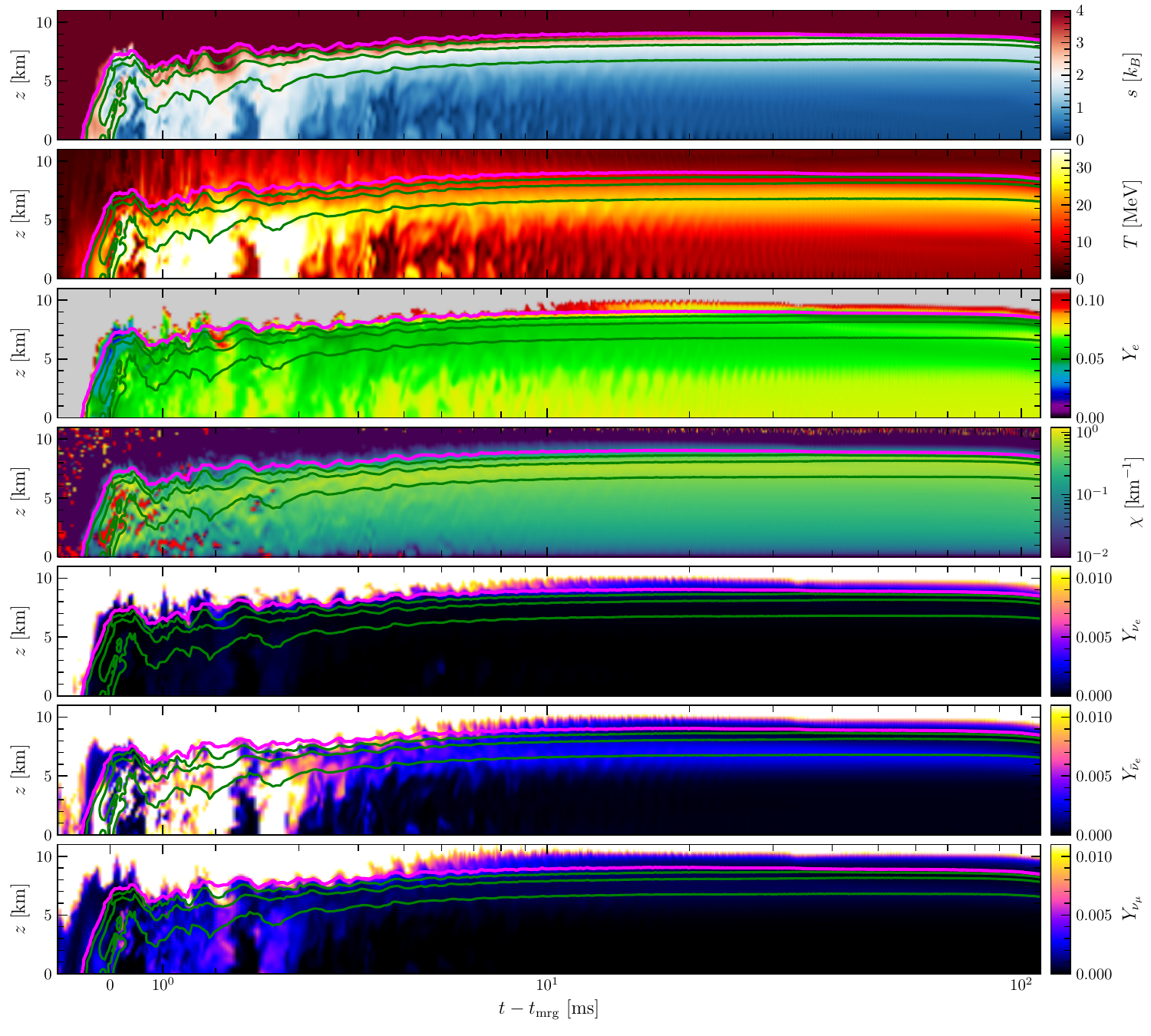}
  \caption{Profiles of entropy per baryon $s$, electron fraction $Y_e$,
  convection instability criterion $\chi$, and neutrino fractions
  $Y_{\nu_e}$, $Y_{\bar{\nu}_e}$, and $Y_{\nu_\mu}$ along the $z$-axis
  for the $\ell_{\rm mix} = 0$ SR binary. The profiles are shown as a
  function of time from merger $t - t_{\rm mrg}$. The purple contour
  denotes $\rho = 10^{13}\ {\rm g}\cdot {\rm cm}^{-3}$, while the green
  contours are $\rho = 10^{13.5}, 10^{14}$, and $10^{14.5}\ {\rm g}\cdot
  {\rm cm}^{-3}$. This figure should be contrasted with
  Fig.~\ref{fig:profiles_xy}, which shows the same quantities in the
  equatorial plane.}
  \label{fig:profiles_z}
\end{figure*}

Figure~\ref{fig:profiles_z} shows space and time profiles of various
thermodynamic quantities along the rotational axis of the $\ell_{\rm
mix}=0$ SR model. This figure should be contrasted with
Fig.~\ref{fig:profiles_xy}, which shows the same quantities in the
equatorial plane. The main difference is that the density scale-height
is smaller in the vertical direction, due to the absence of centrifugal
support in the $z$-direction. As in the equatorial plane, the peak of
the temperature is at $\rho \simeq 10^{14.5}\ {\rm g}\ {\rm cm}^{-3}$,
although the maximum temperature in the vertical direction (${\sim}25\
{\rm MeV}$) is somewhat smaller than the peak temperature on the
$xy$-plane (${\sim}30\ {\rm MeV}$). This is also the region at which
$Y_{\bar\nu_e}$ peaks, $Y_{\bar\nu_e}$ is about a factor two smaller
than in the equatorial plane. The electron fraction $Y_e$ slowly
increases with time in the regions close to the stellar surface, with
$Y_e$ going from 0.06 to 0.08 at $\rho \simeq 10^{13}\ {\rm g}\ {\rm
cm}^{-3}$. Finally, we compute the Ledoux instability criterion $\chi$
using the data along the $z$-axis and find that the \ac{RMNS} is stable
against convection along the rotational axis.

\section{Effect of Viscosity}
\label{sec:appendix:visc}

\begin{figure*}
  \includegraphics[width=\textwidth]{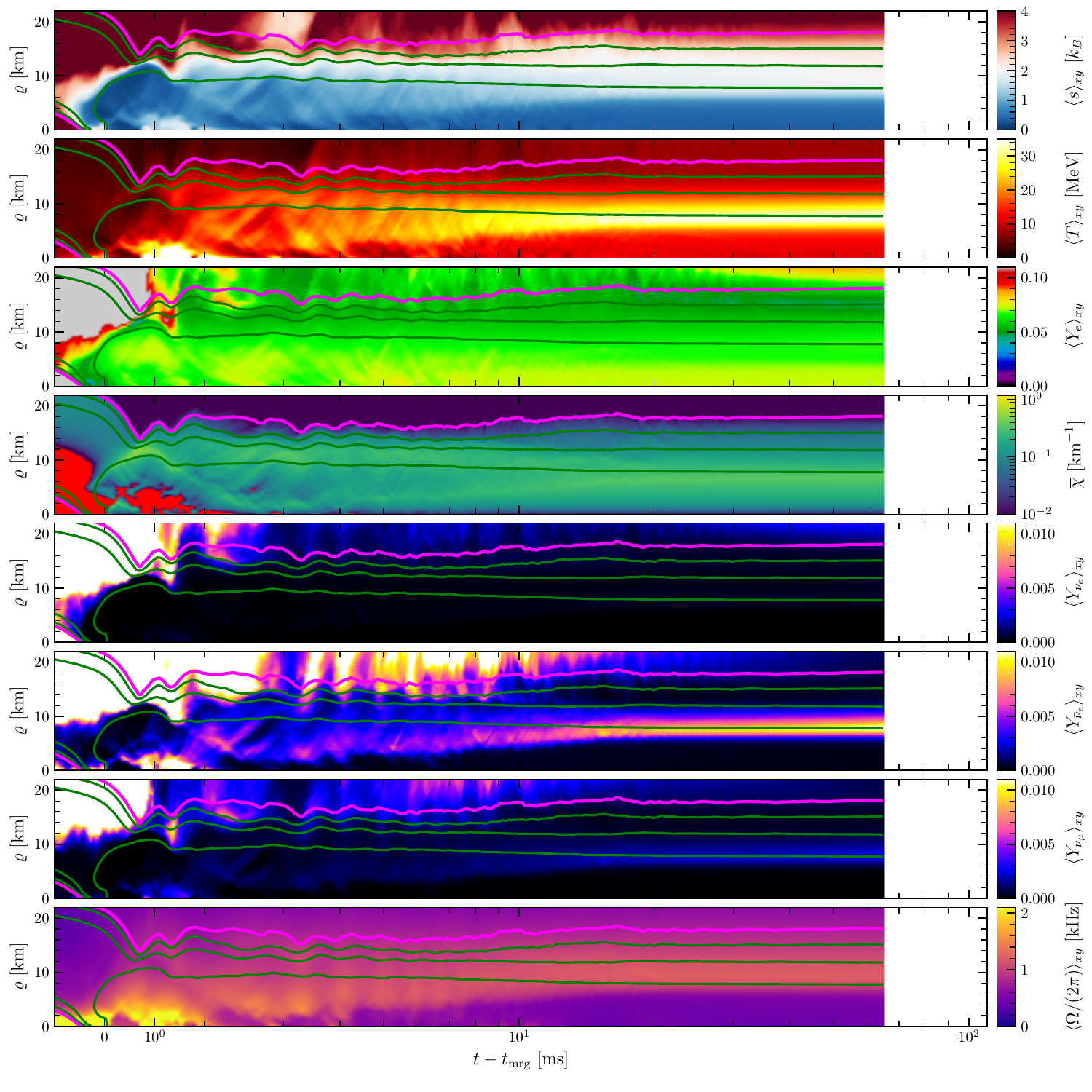}
  \caption{Angularly averaged profiles of entropy per baryon $s$,
  electron fraction $Y_e$, convection instability criterion $\chi$,
  neutrino fractions $Y_{\nu_e}$, $Y_{\bar{\nu}_e}$, and $Y_{\nu_\mu}$,
  and angular frequency $\Omega$ on the equatorial plane for the
  $\ell_{\rm mix}(\rho)$ SR binary. The profiles are shown as a function
  of cylindrical radius $\varrho$ and time from merger $t - t_{\rm
  mrg}$.  The purple contour denotes $\rho = 10^{13}\ {\rm g}\cdot {\rm
  cm}^{-3}$, while the green contours are $\rho = 10^{13.5}, 10^{14}$,
  and $10^{14.5}\ {\rm g}\cdot {\rm cm}^{-3}$. This figure should be
  contrasted with Fig.~\ref{fig:profiles_xy}, which shows the same
  quantities for the $\ell_{\rm mix} = 0$ binary.}
  \label{fig:profiles_xy_k2}
\end{figure*}

Figure~\ref{fig:profiles_xy_k2} shows azimuthally averaged space and
time profiles for several thermodynamic quantities in the equatorial for
the $\ell_{\rm mix}(\rho)$ SR binary. This figure should be contrasted
with Fig.~\ref{fig:profiles_xy}, which shows the same quantities for the
$\ell_{\rm mix} = 0$ binary. The main effect of the viscosity is to
suppress small scale oscillations of the remnant, as can be seen from
the lack of oscillations in the $\rho = 10^{13}\ {\rm g}\ {\rm cm}^{-3}$
surface position (marked in purple in the figure), and to increase the
entropy in the outer core of the \ac{RMNS}. Despite these changes, the
viscous model remains stable against both convection and the \ac{MRI}.
In fact, the increase in the radial entropy gradient due to viscosity
has a stabilizing effect.

\begin{figure}
  \centering
  \includegraphics[width=0.5\columnwidth]{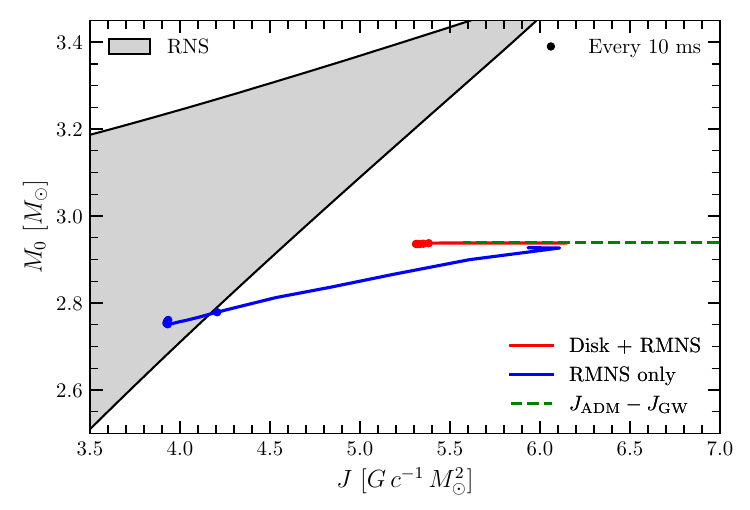}
  \caption{Baryonic mass and angular momentum at the end of the
  simulations for the $\ell_{\rm mix}(\rho)$ SR binary. The grey shaded
  area shows the set of all rigidly-rotating equilibrium configurations.
  The green line shows the evolution of the angular momentum due to
  gravitational wave emission, while the red and blue line show the
  evolution of the Baryonic mass and angular momentum for the disk and
  the remnant, respectively. This figure should be contrasted with
  Fig.~\ref{fig:phase_space}, which shows the same quantities for the
  $\ell_{\rm mix} = 0$ binary.}
  \label{fig:phase_space_k2}
\end{figure}

Figure~\ref{fig:phase_space_k2} shows the evolution of the $\ell_{\rm
mix}(\rho)$ SR binary in the $J{-}M_0$ plane. It should be contrasted to
Fig.~\ref{fig:phase_space}, which shows the same analysis for the
$\ell_{\rm mix} = 0$ SR binary. There are some quantitative differences
between the two binaries, most notably the disk mass for the $\ell_{\rm
mix}(\rho)$ binary is slightly smaller than that of the $\ell_{\rm mix}
= 0$ binary ($0.2\ M_\odot$ vs $0.25\ M_\odot$). The mass loss rate from
the disk is also slightly enhanced with viscosity, however this
difference is within the numerical uncertainty we estimate for our
simulations. That said, the evolution of the $\ell_{\rm mix}(\rho)$
binary is qualitatively identical to that of the $\ell_{\rm mix} = 0$
binary. In both cases, the \ac{RMNS} settles to a quasi-steady
configuration within the phase-space region allowed for rigidly rotating
equilibria, while disk and ejecta take the excess angular momentum.

\clearpage
\section*{Erratum}
\begin{figure}[b]
  \centering
  \includegraphics[width=\textwidth]{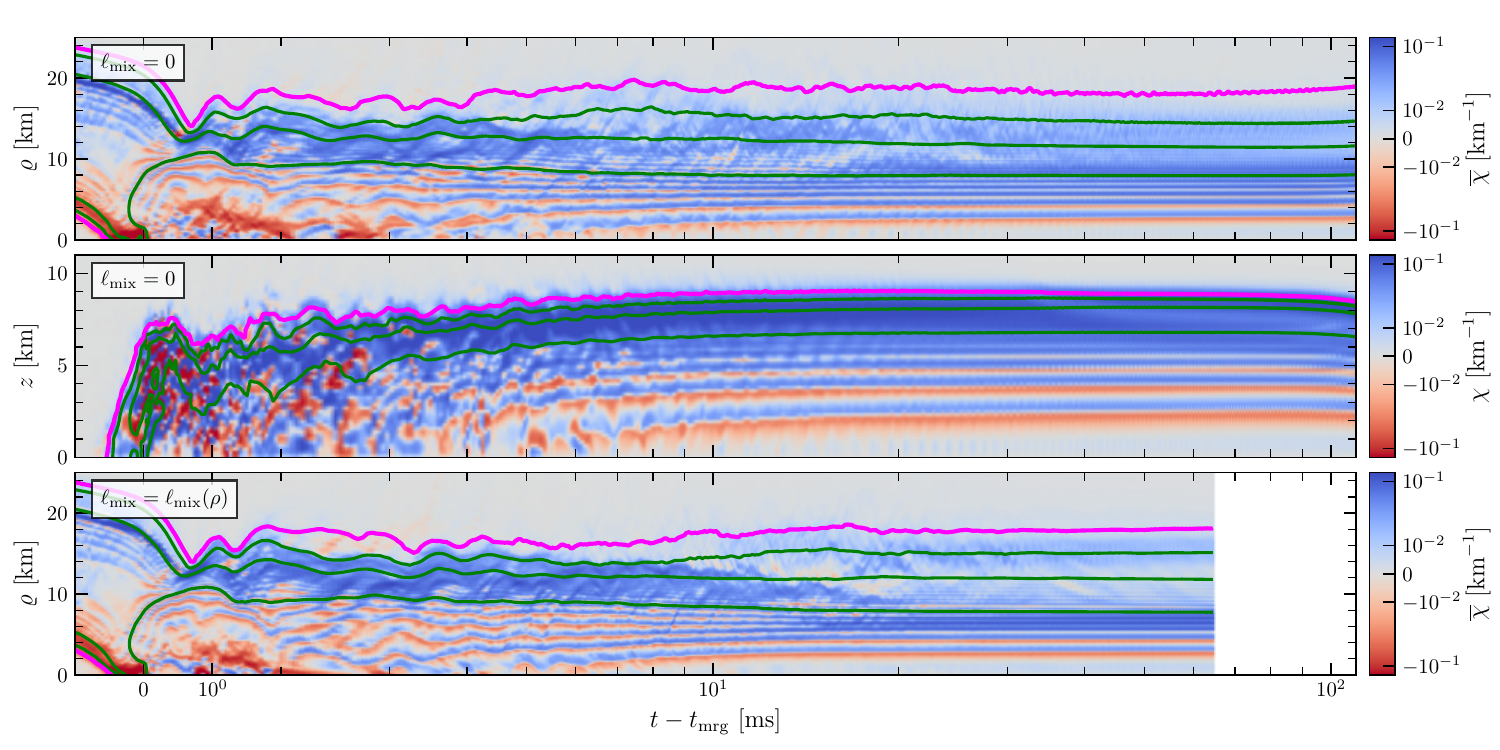}
  \caption{Profile of the convective instability criterion $\chi$ for
  the $\ell_{\rm mix}=0$ and $\ell_{\rm mix}=\ell_{\rm mix}(\rho)$ as a
  function of the cylindrical radius $\varrho$ (averaged over $\phi$ in
  the $z=0$ plane), or along $z$ ($\ell_{\rm mix}=0$ model only).}
  \label{fig:erratum}
\end{figure}

In \citet{Radice:2023zlw}, we presented long-term general-relativistic
neutrino-radiation hydrodynamics evolution of remnant massive neutron
stars (RMNSs) formed in binary neutron star mergers. One of the reported
analysis is the calculation of the quantity:
\begin{equation}\label{eq:chi}
  \bar\chi = \frac{1}{\rho_0} \left[ \left( \frac{\partial \rho}{\partial p} \right)_{s, Y_e}
  \frac{\partial \langle p \rangle_{xy}}{\partial \varrho} - \frac{\partial
  \langle \rho \rangle_{xy}}{\partial \varrho} \right]\,.
\end{equation}
In the previous equation, $\rho$ represents the rest-mass density, $p$
denotes pressure, $s$ indicates entropy, $Y_e$ signifies the proton
fraction, and $\varrho$ refers to the cylindrical radius. The initial
term within the parentheses of Equation~\eqref{eq:chi} is derived from
the equation of state (EOS). A normalizing factor of $\rho_0 = 2.7\times
10^{14}\ {\rm g}\ {\rm cm}^{-3}$ is applied to ensure that $\bar\chi$
possesses units of inverse length. When $\bar\chi$ exceeds zero, fluid
parcels that are displaced upward under adiabatic conditions will have
greater density than the surrounding medium, preventing them from
achieving buoyancy. Put differently, convective stability is maintained
in the fluid when $\bar\chi > 0$. If $\bar\chi < 0$ the fluid might
become convective. However, because this analysis neglects the
stabilizing effect of rotation $\bar\chi < 0$ is a necessary, but not
sufficient condition for convective instability.

We have discovered a bug in the code we used to compute
$(\partial\rho/\partial p)_{s, Y_e}$ in Eq.\eqref{eq:chi}, which
affected the value of $\chi$ reported in the paper. We have recomputed
$\chi$ after having fixed this issue and report the new results in
Fig.~\ref{fig:erratum}. The figure also reports the analogous quantity
$\chi$, which is computed along the rotation axis of the $\ell_{\rm
mix}=0$ binary. While $\bar\chi$ and $\chi$ are positive over most of
the RMNS, consistently with our previously reported results, there are
now isolated regions where $\bar\chi$ and/or $\chi$ are negative.  These
regions have a width of only one grid point, so the sign changes are
likely due to numerical noise in the calculation of the derivatives.
These regions are also located deep inside the remnant, so the
conclusions in \citet{Radice:2023zlw} regarding the role of
stratification in the trapping of the magnetic field are unchanged.


\clearpage

\end{document}